The Cost of Climate Action: Experimental Evidence on the Impact of Climate Information on Charitable Donations to Climate Activism

Samantha Gonsalves Wetherell[1*], Anna Josephson[2]

**Abstract**

We examine the propensity of individuals to donate to climate activism, evaluating the impact of different informational treatments on an incentive compatible charitable donation and stated climate change-related concerns. Participants were evaluated on climate literacy and general climate attitudes before being randomly assigned to a treatment which provided either (1) education or (2) neutral language about climate change, either (3) with or (4) without images of protest. After the treatment, participants engaged in an incentive compatible dictator game. We find that participants gave more to climate activism than seen in previous dictator game and charitable giving experiments, in both average amount given and proportion of participants who gave their entire endowment. However, we determine that climate activism information negatively influenced the amount of money donated. We also found that protest imagery moderated this negative effect and had a positive significant effect of increasing participants' climate concern. Finally, we found that the climate concern was significantly positively correlated with donations, while being a male was significantly negatively associated with donation amounts.

[1] Graduate Student, Smith School of Enterprise and the Environment, School of Geography and the Environment University of Oxford, OUCE South Parks Road Oxford, OX1 3QY United Kingdom of Great Britain and Northern Ireland; [2] Associate Professor, Dept. of Agricultural and Resource Economics, University of Arizona, 650 N. Park Avenue  Tucson, Arizona 85719 United States of America
* Corresponding author.
This work was funded by the Institute of Educational Sciences, the University of Arizona Honors College, and the University of Arizona Graduate and Professional Student Council.  The authors would also like to thank Haydon Eckstrom, Leah Callovini, Michelle Perfect, Brandy Perkl, and everyone at UArizona Divest for their support as well as proofreaders Laura Hernandez and Rhys Williams. Replication package is available on GitHub: https://github.com/aljosephson/climate-change_charitable-behavior.

## 1. Introduction

Climate change has irreversible negative effects on all sectors of life, from biodiversity to infrastructure to health to individual livelihoods. More than 75% of individuals find climate change to be a major threat to their society (Poushter et al., 2022). For United States voters, climate change is a top ten issue when voting for a candidate (Leiserowitz et al., 2021). Addressing climate change is a collective action problem. Climate activism has been shown to be a solution to this problem through changing and strengthening political and public will on climate change action (Swim et al., 2019). Yet, how much people value the effect of climate activism is highly dependent on their beliefs, concerns, and literacy on climate change and activism.

There is a lack of climate literacy among the United States population, including among university students (Bedford, 2015). Low climate literacy can be attributed to a lack of comprehensive environmental education on the issue of anthropogenic climate change, especially the lack of information on the sociological implications of and political, legal, and economic solutions for climate change (Stevenson, 2007). However, even among those who do have a more comprehensive understanding of climate change and the importance of climate activism, direct action efforts to mitigate the effects of climate change are still limited. While climate activism can range from actions such as contacting a representative and attending a protest to donating money and volunteering for an activity that is focused on combatting climate change, only 24% of United States adults reported performing any of these actions in the last year as of 2021 (Tyson et al., 2021). Not all climate activism efforts are viewed equally either. Whether it be notions on the effect of an individual participating in a collective action movement (Kutlaca et al., 2020) or perceptions of different modes of activism, many people may have strong feelings for or against certain kinds of climate activism, especially the very traditional forms of protest like marches or civil disobedience (Bugden, 2020). In fact, among all United States adults who reported making efforts to support climate action, sending donations was the most common method of support while protesting was the least common method of support for climate action (Tyson et al., 2021).

To determine people's propensity towards climate activism, specifically, their donation habits, we use a modified dictator game (generally following Shreedhar and Mourato (2019)). In our experiment, participants are evaluated on their climate literacy and general climate attitudes before they are randomly assigned to one of four informational infographic treatments. The treatments are either an infographic, informational treatment with neutral climate change information (1) with or (2) without additional information on a climate activism organization and (3) with or (4) without an image of protest. After the treatment, participants' climate attitudes are reassessed, and they participate in a modified dictator game in which they choose whether to donate to a climate activism organization from an allocation of $30. Based on this, we assess how treatments change participants' propensity to donate to climate activism groups and examine how these materials affect their climate change attitudes.

From our experiment, we find that the average donation is around half of the endowment ($14.28). This exceeds regular dictator game donations two-fold and is at the upper-bound for charitable giving experiments. Moreover, we find that people were twice as likely to donate their entire endowment than to keep the entire endowment; this is the opposite of what is found in other dictator games (Engel, 2011) and charitable giving experiments (Shreedhar & Mourato, 2019). Additionally, we find that information treatments have a significant negative effect on donation amount, while identifying with a gender other than cisgender male, what we refer to as a non-male person for simplicity, and having climate concern have a significant positive effect. Finally, we discover that the protest only treatment had a significant positive effect on climate concern.

The driving factor behind the overall high donation results is the equally high concern for the issue of climate change, motivating a positive valuation of climate activism as a problem-solving tool by our participants. We also believe that our negative information treatment effect may be a result of the free-riding effect as seen amongst political campaign supporters who exerted less effort upon learning that peer supporters planned to exert more effort than expected for a campaign in Hager et al. (2020). Upon learning that a large, organized group is combating climate change, people become less concerned about the need to contribute to the cause of climate activism. Yet, conversely, images of grass-root protest dampen this effect by increasing climate concern once seeing people perform activism.

There are numerous studies on the willingness to pay for climate change adaptation measures. Yet, there is little information on social valuation measures beyond willingness to pay and how much people will fund the sociopolitical mechanisms needed to bring about these often-technical solutions, such as through the mechanism of climate activism. Moreover, there is little information on how these funding preferences for climate activism may be affected by knowledge and opinions on climate change, climate activism, and protests in climate activism. In this study, we investigate a new aspect of funding sociopolitical climate change solutions by eliciting donation preferences for climate activism rather than willingness to pay (WTP) for climate adaptation (Mayer & Smith, 2018; O'Garra & Mourato, 2015; Vilela et al., 2022). We examine donation preferences to assess how different informational interventions may affect people's attitudes towards climate activism and their donation preferences associated with climate activism.

This study is the first of its kind with respect to climate solution valuation and opens doors to understanding people's donation habits towards climate activism as well as how those habits coincide with their attitudes about climate change and activism. Through this experiment we can understand how to engage various populations on the issue of climate change and how information may either motivate or deter individuals to contribute to climate change solutions. This research presents a new economic lens on the issue of climate activism and climate change education. The experiment's design shows how environmental and behavioral economists can look to sociopolitical solutions to climate change, beyond the technical. For climate activism organizations, our research provides useful evidence as to how to entice donors who have varying levels of climate concern. Overall, our experiment demonstrates to people on a broader level that activism has political, social, and monetary value.

We begin with our motivation and a review of previous literature. We first delve into different non-market valuation methods and experiments followed by a discussion of the value of collective action and activism as well as people's perception of forms of activism. We then move on to our experimental method where we discuss how we determine the differences in climate knowledge, climate attitudes, and individual donation behavior using a climate literacy test, pre-treatment and post-treatment attitudes tests, and a dictator game. Moreover, we detail our infographic-style informational treatment, with a cross-treatment which integrated images of protest into the informational treatment to determine the effects of media content of climate activism/protest affected climate attitudes and donation preferences. Afterwards, we describe our data, discussing our participants, variables, and our method of empirical analysis. From there, we then go deeper into our analysis and its meaning in our summary statistics section. In this part, we cover our summary statistics and our preferred regression analyses. We conclude with our discussion and conclusion section where we summarize our findings, discuss their implications, and give our recommendations based off them.

## 2. Motivation & Literature Review

In the following section, we review literature around climate valuation, activism, and climate literacy. We begin by reviewing different climate valuation methods and their efficacy, and then move into how the tool of political activism operates in climate change solution making as well as the effect of protest on public perceptions of a movement. We finish by examining how the intermingling of people's perception of climate change and activism in conjunction with different levels of climate literacy effects people's climate beliefs.

### 2.1 Valuation Methods

The nonmarket valuation measure of consumer willingness to pay (WTP) has been an essential tool for economists to determine the benefits and desire for climate change mitigation and adaptation solutions. Compared to external costs and benefits assessments that require extensive and varied data, WTP demonstrates social valuation by eliciting revealed and/or stated preferences from individual consumers. Yet, there is wide variation in results of stated WTP for climate change solutions and difficulty in ensuring that people's external economic decisions reveal certain preferences on climate change and are not a consequence of other factors in the results of revealed WTP. Chaikumbung (2023) dissects these major variations in WTP by determining the main drivers of WTP and the effect of cultures and institutions on preferences and WTP for climate change solutions, finding that factors related to institutions, cultures, individual attributes and environmental attributes affected WTP. With this variation in WTP and the tendency for high WTP estimates that do not correspond with people's income (Murphy et al., 2005), researchers have recently explored new ways to estimate social valuation through individuals. Shreedhar and Mourato (2019) address this issue in their dictator game experiment by using donation amounts towards a conservation organization to determine how people's pro-social behavior towards habitat protection is affected by audio-visual media. The authors identify a relationship between the use of charismatic species and increased probability of donating and a link between information on anthropogenic climate change's effect and an increased margin of donations without the use of WTP. This experimental structure suggests that dictator games may provide a clear and robust manner of estimating social valuation that is more applicable to the real world as opposed to traditional WTP.

## 2.2 Collective Action, Activism, and Perceptions of Protest

While many studies highlight the importance of collective action in bringing about the collaborative problem-solving needed for managing public goods like the climate (Bernard et al., 2013; Ostrom, 2000; Walker et al., 2000), none of these studies discuss the issue of how to bring about collective action or the valuation of it. Defining collective action, we find that collective action is any action that is done by a group to achieve the goals of that group. Accordingly, activism can be defined as collective action performed for the purpose of creating social change and/or political change. Reviewing collective action literature, we uncover that the use of protest has been a significant driver in building collective action and creating a force for change in various movements (Moyer, 1987). Swim et al. (2019) delves into how climate protest impacts bystanders' efficacy beliefs, perceived social norms, perceptions about public climate change concerns, impressions of protestors, subsequent behaviors, and whether these impacts are different based on "the political leanings of the bystanders preferred news source." The authors reveal that marches increase perceptions of collective efficacy on climate change issues and decrease negative perceptions of marchers, indicating the effectiveness of climate protest in building the basis for collective action in climate change solutions. Further, climate activism may not only increase beliefs in efficacy but also, specifically, contribute to more support for pro-environmental policy. Kountouris and Williams (2023) examine the influence of climate protest on the public's environmental attitudes, focusing on perceptions of sustainable lifestyles, their own environmental impact, pro-environmental policy and behavior, and the severity and imminence of the climate crisis. They identify a significant increase in approval of pro-environmental policies and behavior post-protests, a decrease in willingness to pay a premium for environmentally friendly goods after the protests, and no negative environmental sentiment backlash from the protests. From these studies, climate activism is shown to be an essential tool in influencing public and political sentiment on climate policy and creating an environment for the collective action needed to enact climate change mitigation and adaptation.

However, not all forms of protest are viewed equally, nor are all people equipped with the knowledge to understand the importance of climate activism generally. Bugden (2020) investigates the effect of three types of climate protest (marches, peaceful civil disobedience, and violent protest) on support for changes in the structure of society or "sentiment pools" and how these effects are constrained by partisanship. Bugden (2020) finds that all parties, regardless of experimental condition, are relatively equally supportive of climate protest but that civil disobedience and, more so, marches increase sentiment pools the most for Democrats and Independents who believed in anthropogenic climate change with no effect on Republicans. Moreover, there is a small backfire effect from Independents who do not believe in anthropogenic climate change in response to civil disobedience and violent protest. Therefore, while the most well-received protest amongst the population is that of peaceful marches, there are clear differences in sentiments towards climate activism between people who believe in or do not believe in anthropogenic climate change. This suggests that people's lack of climate literacy may hinder their ability to understand the motivations behind or support such movements, which could pose a problem in helping to organize collective action. Bedford (2015) uses a case study to examine the relationship between climate literacy and concern for anthropogenic climate change among higher education students. Bedford finds that climate change literacy and attitudes towards climate change were mainly segregated on party lines. However, the study also found that increased climate literacy decreased polarization between Republicans and Democrats and increased climate change concern for Republicans. Thus, while Republicans trail behind in climate literacy, we can find that, overall, increased information on climate change can assist in literacy and help individuals recognize the need for climate activism, independent of party affiliation.

## 3. Experimental Method

We examine the differences in individual donation behavior and differences in climate knowledge and concern after participants are randomly assigned to view an infographic-style informational treatment, with a cross-treatment which integrated images of protest into the informational treatment. We use a between-subjects approach to examine the effects of media content surrounding climate activism/protest on donation preferences. For a visualization of our method, please refer to Figure A1 in Appendix A.

To measure the outcome of these treatments on donation behavior and perception of climate change and action, we use an incentivized dictator game, and employ tests (pre- and post-treatment) on climate attitudes.

### 3.1 Climate Literacy Test

To examine variations in climate literacy, we employ a test of climate literacy at the very beginning of our experiment. We use a survey modeled on the US Global Change Research Program's (USGCRP, 2009) and Bugden (2020) climate literacy survey. The instrument is available in Appendix B. Based on the number of questions correctly answered, we then determine whether someone is climate literate, with a score of 70% or more qualifying as climate literate. This test was only given before the treatment. After completing it, participants moved on to the climate attitudes survey.

### 3.2 Climate Attitudes Test

For our climate attitudes survey, we use questions from the Yale Program on Climate Change Communication national survey as it is considered the standard for evaluating climate change attitudes and perceptions (Leiserowitz et al., 2021). We partitioned the test into two parts to ascertain (1) an individual's belief in the existence of climate change, (2) confidence in that belief, and (3) their level of concern on the issue of climate change. This test was deployed both before and after the infographic treatment. This instrument is available in Appendix B.

### 3.3 Experimental Game

When looking for a climate activism organization to use in our experiment, we searched for an organization that was nonpartisan, relatively well-established, multifaceted in their climate activism engagement approaches (e.g., performed protests, engaged on social media, and lobbied elected officials), and clear in their organization's goals. From these criteria, we chose the organization Citizen's Climate Lobby (CCL) due to its stance as a nonpartisan, nonprofit climate change organization that performs grassroots, organizing everyday people, to grasstops, lobbying national legislators, advocacy to use in our infographic-style education treatments and our dictator game.

To investigate individuals' propensity to donate to CCL, given their status as a climate activism organization, we create four infographic treatments with neutral climate change information (1) with or (2) without additional information on a climate activism organization and (3) with or (4) without an image of protest. To elaborate, we use a four-armed treatment and control structure: we first have an education treatment in which an education treated individual receives information about both climate change and climate activism and an education untreated (control) individual just receives generic information on climate change. Within those groups, there is an additional protest treatment, in which individuals who are protest treated receive (within their education treatment or control) images of protest and individuals who are not protest treated receive neutral images of people tabling (A.2 in Appendix A). The participants were randomized with equal probability into one of the four treatment groups automatically via Qualtrics after completing the climate literacy test and pre-test climate attitudes survey. In all treatments, we featured three small paragraphs of information on climate change. This information was neutral and defined climate, climate change, and the effects of climate change. This information was made following the same US Global Change Research Program document on climate literacy (USGCRP, 2009) that was used to construct our climate literacy test. For the two of our infographics that include information on a climate activism organization, we added an additional small paragraph about the organization CCL. This paragraph gave information about the organization and its work for climate activism so we could measure the impact of information about a climate activism organization on participant's donations and attitudes. This snippet highlighted the organization's nonpartisan efforts, its grassroots to grasstops approach, and its multifaceted manner of advocating for climate change solutions. The formatting for all infographics was standardized. All infographics had the same color scheme, font, and have the inclusion of an image at the bottom. In the neutral information and neutral plus climate activism information group, this image will either be an image of CCL group members gathering while tabling or be an image of CCL group members at a rally smiling. All treatments can be found in Appendix A.

After receiving the treatment and completing the post-test climate attitudes survey, participants participate in a dictator game. The participant plays the role of the "dictator" and is asked to decide how to allocate an endowment of $30 between themselves and the "recipient" (CCL). This allocation was done on a donation sliding scale of $1 integers, with the default being set to $15. The participants could allocate any amount they desire to themselves and to CCL, allowing them to

donate nothing or everything. The participants were told that they would be able to keep the initial endowment minus however much they decide to donate.[1] As with treatments, all donation pages were standardized. Every donation page displayed a photo of the CCL logo along with one short sentence describing CCL as "a climate activism organization that works to bring about public and political will for climate change solutions." By having this standardized donation page, we hoped to further bridge the gap between our experiment and real-life charitable giving scenarios by following the standard format of climate activism donation pages. Furthermore, by including the same snippet of information for every participant on CCL, we allowed all participants to understand the basic function of CCL without giving them in-depth information on the organization, as done in the Climate Activism Information treatment.

While many dictator games give participants an initial endowment of $10 (Engel, 2011), we follow Shreedhar and Mourato (2019) and use an endowment of £25($30). This amount is relatively high stakes in comparison to the participation fee, which could dampen charitable behavior. However, we believe this amount will most closely replicate real-life charitable giving situations.

Following the dictator game, participants respond to a series of questions on environmental and pro-social behavior. For determining pro-environmental behavior, we developed a measure following Mateer et al. (2022). Each question in the test is rated on a 5-point Likert scale from "Never" to "Always" for participants to answer how often they perform different pro-environmental behaviors. However, those who have previous pro-social behaviors and have contributed to other activist causes unrelated to the environment are not considered in these questions. As younger donors are more likely than older donors to engage in advocacy for an organization or cause (Dorothy A. Johnson Center, 2024) and 70% would rather donate time than money to a cause (Canagarajah, 2021), this subset of non-environmentally engaged volunteers is important to capture. Thus, we add an additional four questions in the same style of the pro-environmental behavior (PEB) questions to capture general pro-social behavior (PSB).

Finally, participants respond to a brief set of questions about individual attributes including age, race, gender, and work status. All surveys can be seen in Appendix B.

---

[1] One of every 10 participants were randomly selected to receive their actual payout from the dictator game, along with the $5 dollar participation fee received by all participants. All participants were informed that every participant has an equal and fair chance of receiving said payout in the consent form before beginning the experiment.

## 4. Data

In this section, we present the data used in our analysis, including summary statistics on our participants and details on our independent variables and control covariates.

### 4.1 Participants

We recruited 367 students from a large, R1 university in the western United States through email recruitment lists. The experiment ran from October 19 to December 19, 2023, through the Qualtrics survey platform. All subjects were informed that the survey was on climate solutions and attitudes and was open only to undergraduate students at the university. Our focus on a student population presents us with a lower bound of observed effect size as students are likely to donate less than other populations in dictator games (Shreedhar & Mourato, 2019).

In our sample, 64% are women and 31% are men. The average age of a participant was between 18 to 24 years old and 53% of the sample were full time students only with the rest being partially or fully employed while being a student. Of our sample, only 13% say they never donate to charity while most donate either rarely (33%) or sometimes (40%). 42% of participants report they have never donated to support local environmental protection.

### 4.2 Dependent Variables of Interest

In our experiment we used three dependent variables: donation amount (see Figures 1 – 4), post-treatment climate concern, and post-treatment climate assuredness. Of these three variables, two of them, post-treatment climate concern and climate assuredness, come from the post-treatment climate attitudes survey. The purpose of these post-treatment climate attitude variables was to measure how the infographic treatment altered people's assuredness in the existence of climate change and their concern about climate change. On the other hand, the purpose of the variable of donation amount was to measure how information treatments altered people's charitable giving behavior towards a climate activism organization, or to determine how much they valued a climate activism organization's contribution in climate change solution making.

### 4.3 Independent Variables

We also measured several independent variables before and after the treatment. Before our treatment, we gathered information on individuals' climate literacy and on pre-treatment climate attitudes. Pre-treatment climate attitudes are divided into climate change concern and assuredness. After our treatment and our subsequent attitudes survey and dictator game, we assessed previous pro-environmental behavior, inclusive of environmentally conscious habits such as limiting energy usage all the way to participating in an environmental activism group, and pro-social behavior. Finally, we gathered demographic data and information on people's political leanings, which was determined using a sliding scale that went from zero (the furthest left) to 100 (the furthest right). All variables are presented in Table 1.

**4.4 Empirical Analysis**

We used a between-subjects design with exposure to multiple treatments with information and exposure to cross-cutting treatment with protest imagery. Our random assignment of subjects into treatment groups controls heterogeneity in the subject pool, along with our control of the individual attribute variables.

Before our regressions, we examined correlations to determine which variables had a statistically significant relationship with one another. Afterwards, to control for heterogeneity in participants, we selected the covariates of participation in climate solutions, gender, political orientation, climate literacy, and the level of climate concern. In our regressions, we were interested in three treatments. First, we have the treatment variable of images of CCL gathering versus images of protest. Secondly, we will have the treatment variable of climate activism information. Thirdly, we will have the combined protest imagery treatment and climate activism information treatment.

We then conducted our first analysis: a censored model of a Tobit regression. Donations are measured as a continuous variable between 0 and 30. Afterwards, we used an ordered logit regression to further explore donation behavior, with donations binned into eight different donation groups. For ease of understanding, I will continue to refer to the donation amounts in these bins in number format and the bin number in word format. Seven of these bins were split up into intervals of five for the integers 1 through 30, with the smallest bin, bin one, being 0 alone. From this, we analyzed the treatment effect on the likelihood of people donating within different categorical bins.

Finally, to explore the effect of information on climate attitudes, we again used a Tobit regression, censored at the minimum, 5, and maximum, 21, climate concern scores.

## 5. Results

In this section, we discuss our findings. The section begins with a review of summary statistics and the examination of some general relationships between our variables of interest and independent variables. We then present our findings from our econometric specifications. The section concludes with a discussion of the findings.

### 5.1 Summary Statistics

We find that the average donation was $14.28 or 47.6% of the allocation. We also find that 86.92% of participants donated, with 16.35% of participants donating all their endowment, while 13.08% donated nothing. The median amount given was $15 (18.26% of the sample). The amount donated is higher than expected, though coincides with upper limit of the average donation in previous charitable giving experiments, which ranges from 30% to 50% of the allocation, (Cartwright & Thompson, 2022) though contrasts with the average dictator game donate, which is around 20% of the allocation (Levitt & List, 2007). Yet, what does not coincide with the literature, is how many people gave their entire endowment. In comparison to normal dictator games, in which around 5% of the sample give their endowment (Engel, 2011), or even a charitable giving experiment, where donations seem to be slightly higher at around 6% (Shreedhar & Mourato, 2019), our experiment had over three times that amount give their entire endowment. Figure 2 illustrates the donation amounts from all participants in a histogram. In this histogram, we can also see clustering of donations around intervals of $5, indicating that more people donate in round figures. This follows the natural human behavior to donate in round number intervals, due to round number bias (Coupland, 2011).

Next, we consider donations by treatment groups. This is presented in Figure 2. For those who received the information treatment, donations averaged at $13.38, which was lower than those who did not receive information ($15.17) and lower than that of those who did receive ($14.15) and did not receive ($14.42) the protest imagery treatment. Yet, when looking into smaller subsets of the sample, we can see how the effect of the treatments differs for those who have different climate concern. Notably, while information has a negative effect for the entire population, the addition of protest as a treatment seems to raise the amount of money donated for those with lower climate concern and decrease the amount donated for those with higher climate concern as seen in Figure 3 and Figure 4.

*Tobit: Donation*

In Table 2, we present the results of the Tobit model. We use a Tobit model as it is a censored regression model that allows both lower and upper limits, which we have in the form of 0 and 30. The results support the idea of the negative effects of the information treatment, as shown in our previous figures. We see that individuals exposed to the information treatment were less likely to donate, relative to those not exposed to the information treatment. In model 1 and 2, we can see that the information treatment is associated with $4.90 lower donations on average. The combined information protest treatment only leads to about $3.60 lower donations. Yet, the protest treatment does not lead to significantly less donations. This finding provides tentative evidence that climate activism information lessens people's donations while protest imagery has no significant effect on them. However, if one was exposed to the protest treatment and the climate activism information treatment, it seems the negative effect from the information treatment is somewhat attenuated, with both the significance and magnitude of the negative effect of treatment lessening.

Additionally, we can see that concern and being non-male had a significance of 1% and led to around $13 and $5 higher donations respectively. This result coincides with previous literature which demonstrates that women are more likely to donate and, when they donate, donate more on average (Mesch et al., 2011, Piper and Schnepf, 2008) as well as that people who are more concerned with climate change tend to donate more to environmental organizations (Shrum, 2021; Zaval et al., 2015). We also observe that in Model 1 or Table 2, the variables of participation in climate change solution making have a positive effect on donations and both political leanings and climate literacy have a negative effect, though none of the effects are statistically significant. This finding of climate change solution making and pro-environmental behavior not having a significant effect on donations differs from previous findings which show that information and social norms can increase pro-environmental behavior (Buckley, 2020).

*Ordered Logit: Donation*

Finally, we consider our findings from the Ordered Logit, which is presented in Table 3. We chose to use an Ordered Logit to be able to see not just the overall effects of the independent variables on donation, but on different donation categories. Our donations are binned into eight different donation groups. Seven of these bins were split up into intervals of five for the integers 1 through 30, with the smallest bin, bin one, being 0 alone.

In this regression specification, we see the same significance and direction for concern, non-male, information, and information protest as we did in the Tobit model. The order of the effect size from each variable, from the largest coefficient from climate concern to the smallest coefficient from information protest, follows the same pattern as the Tobit model too. More specifically, percent of climate concern had the largest effect on donations, followed by non-male, the information treatment, and the combined information protest treatment. With this specification, we observe the least and most probable bins that the donators choose – that is, what amount of money participants were most and least likely to donate. We find that the bin with the highest probability of being chosen is the largest donation bin and the bin with the least probability of being chosen is the very first bin. In fact, we find that people are twice as likely to donate everything over nothing. Delving into this model further, we look to the marginal effect of each variable on the probability of donations being in different bins (Table 3 and 4). With this table, we see that every variable loses significance or becomes insignificant at bin four. This may be caused by an anchoring effect as our default start was placed at $15, which lies within bin four. Moreover, we can also see an increase in the probability of a donation being in bin three for all three treatments and an increase in significance of a negative likelihood of being in bin seven for both information treatments. This suggests that the treatments might have a centralizing or grounding effect on donations, bringing donation amounts closer to the middle bins. We discuss this further in the subsequent section.

*Tobit: Attitudes*

We chose to then use a Tobit model again (Table 5) for our climate concern due to our climate concern score having a lower limit of 5 (1 point for each climate concern question, with an upper limit of 21). Returning to this specification, we can examine the relationship between treatment and participants after treatment climate concern score. First, we can see that pre-treatment climate concern is largest, with everyone additional unit or point of concern translating to an increase of 0.94 of a unit in the final climate concern score as noted by the coefficient of 0.944. Additionally, we can see that pro-environmental behavior has a coefficient of 0.138 and that climate literacy has a coefficient of -0.043 or a decrease of almost half a unit of climate concern. Therefore, the more initially concerned about climate change and the more pro-environmental behavior, the higher the final climate concern, but the higher the climate literacy the lower the climate concern. Moreover, we find that our protest treatment had a significant effect on climate concern. This suggests that seeing images of protest by themselves increases people's climate concern, but that concern is dampened when information is given. This is line with previous research which demonstrates that information on environmental issues or climate issues has an insignificant effect on people's climate concerns (Kountouris, 2022).

*Robustness Checks and Limitations*

The results of our regressions are robust to a variety of models, demonstrating the same significance, weight, and direction. We replicated our models using Cragg-Hurdle and Poisson regression models. Moreover, we checked for heterogenous treatment effects for those of different climate concern, literacy, pro-environmental behavior, gender, and politics for each of the treatments, and found no difference in our results.

Despite our consistent findings, our study is not without its limitations. While we have tried to control for as many factors as possible and to make our information treatments neutral, it is still possible that prior feelings around climate information, climate activism, and protest affected participant decision making. We attempted to control for this by determining of people's climate concern, past pro-environmental behavior, and past pro-social behavior in our surveys, but we can still not fully rule out other factors, such as trust of charities, feelings on non-profits, and sentiments towards different kinds of activism (Shreedhar & Mourato, 2019).

Moreover, we are unable to rule out possible effects associated with our population. Previous work has found that certain populations, such as western, educated, industrialized, rich, and democratic or "WEIRD" (Henrich et al., 2010) people or students, may have behavior that cannot be extrapolated to the larger population. This may provide further opportunity for exploration via replicating our experiment with a sample more representative of the United States population or through a non-student sample. Yet, while students have sometimes been found to be less generous than the overall population (Engel, 2011; Falk et al., 2013) or more generous than the overall population (Carpenter et al., 2005), they also have been found to make similar contributions to the overall population (Exadaktylos et al., 2013). Examining our own student participant sample, it seems that our participants donate more in line with a non-student population than a student one (Belot et al., 2015). Thus, our student donations may either be a good lower bound for donation amounts or comparable to donations for the larger population. However, it should be noted that younger adults have a much higher concern for climate change than older adults, suggesting that these large donation amounts, and lack of no donations is due to the high importance of climate activism to Gen-Z (Funk, 2021).

## 6. Discussion and Conclusion

In this paper, we examined the effect of short informational treatments and images on charitable donation preferences towards climate activism and attitudes towards climate change. In doing so, we delved into a new area of valuing climate change solutions, looking to sociopolitical solutions. Moreover, we attempted to determine how altering an individual's climate concern might affect their charitable donation preferences towards climate. Our results demonstrate that people highly value climate activism as a climate change solution; that information on activist groups causes people to donate less; and people become more concerned about climate change when witnessing images of protest.

Examining first our findings around the relatively large average donation amounts to climate activism and comparing our work with other dictator games, we see that in most dictator games generally have around a 60% positive donation rate and an average of around 20% of endowment donated (Levitt & List, 2007). In charitable giving games, that amount raises to around 30% to 50% of the endowment donated if the subject is a charity (Cartwright & Thompson, 2022). As our average donation of $14.28 or 47.6% of the endowment remains firmly on the upper bound of charitable donation averages, we determine that people value the solution of climate activism above giving to other individual participants and other kinds of charitable organizations. This is

perpetuated by the fact that we observe three times the normal proportion of participants donating their entire endowment and that participants were twice as likely to donate everything than nothing (Engel, 2011; Shreedhar & Mourato; 2019). We can thus conclude that our high average donations and high rate of positive donations are reflective of the high regard that our participants have for climate activism as a solution to climate change.

We also found that people seemed to donate less when exposed to the information treatment. This might be due to several different reasons. First, it may be due to the perceived size of the charity. It has been suggested that Generation Z prefer local or grass-roots movements over larger organizations (Mercado, 2023). This may be further explored in future work which could investigate charitable donation differences to larger compared with smaller activism groups or national compared with local activism groups. However, the opposite has also been found, as people have been discovered to prefer giving to more popular or larger charities over smaller ones (Bachke et al., 2014). So, we consider a second explanation: people donate less to climate activism when they are exposed to information on climate activism organizations due to a free-rider effect. We believe this is the case. We see that concern for climate change is unaffected by either information treatment, yet when all else is held equal individuals donate less. This suggests that participants feel that their donation is not needed because they assume others will donate or contribute to climate activism. This supports past literature that examines political activism as a public goods game with incentives to free ride (Olson, 1965), and field experiments on activism that demonstrate how that effect plays out in real life when one member of a party's campaign demonstrates they'll exert more effort the other members will intend to exert less (Hager et al., 2020). However, we observe that people seem to become more concerned when seeing images of protest alone and that our observed free-riding effect is somewhat dampened when the information and protest treatments are interacted. While this might seem contradictory to this free-riding hypothesis, it is in line with our findings: in seeing images of people protesting, that made the climate activism organization seem more vulnerable and less empowered, thus lessening the feeling that other people had the issue of climate change "handled". This finding aligns with past literature on charitable giving that demonstrates people prefer to donate to projects that help those who seem vulnerable or poor (Cartwright & Thompson, 2022). Yet, it should also be noted, that the free riding of some may not affect the donation amounts of those who chose to donate. In an experiment from Frings et al. (2023), it was demonstrated that individual's WTP for conservation was not affected by the fact that others may decide to free-ride off their donations. Therefore, making it known that donations are needed for an organization and the importance of an individual donor's contribution will likely increase the possibility of donation and the amount donation.

We also have a notable grounding effect around certain donation amounts. People have strong pre-conceived desires to donate certain amounts, including 0, 15, and 30. Most people start at one of these values and then may sway towards another value. We explored the effect of having an information treatment or information protest treatment on the probability of ending up in a certain bin and found that both treatments ended up decreasing the probability that donations would be at the minimum or maximum. Thus, it seems that donations centralized with the treatment, furthering the idea that people start off with wanting to be in donation bin one, four, or seven ($0, $11-15, or $26-30). Moreover, we find that those who want to donate in the middle or bin four ($11-15) will always donate in bin four, as bin four experiences the least amount of significance from every variable that could affect it. When soliciting donations, it may be helpful to allow people the options of donating in $5 increments, as they would naturally gravitate towards them.

Finally, in line with many previous dictator game studies and studies on charitable donations, women give significantly more than men (Engel, 2011; Mesch et al., 2011, Piper and Schnepf, 2008). In fact, not only is there evidence that women give more than men, but evidence that young male students are a uniquely 'selfish' population when it comes to charitable giving (Cartwright & Thompson, 2022). Moreover, when it comes to specifically donating to environmental causes, women were also found to be more likely donators compared with men (Israel, 2007). Thus, our finding that non-male individuals donate more often than men and in larger amounts, is in line with previous research. Still, it should be noted that, while there was a notable difference in donation amounts, there was no significant effect from gender on one's climate concern. This suggests that the gender-based differences in donation amounts are not a result of differing concern, but other differing incentives to donate. As suggested in previous literature, this may be due to different motivations found for men and women who donate, with men being more likely to donate due to self-interest (IUPUI Women's Philanthropy Institute, 2015).

Overall, our results indicate that individuals do value climate activism as a solution to climate change, and value it highly in comparison to donating to other people or causes. Our findings also indicate that climate activism organizations should focus promotional content on the issue of climate change and the need for and the impact of an individual's contribution on combatting climate change to incentivize donations. However, in the case in which an organization does include information about itself and its work, the organization should minimize the amount of information on itself and include photos that emphasize local and grassroots efforts. In emphasizing the importance of an individual's contribution and an organization's local efforts, organizations can avoid the free-rider effect while also appealing to Gen-Z donors. Additionally, while not covered in the scope of this paper, organizations may also offer non-pecuniary incentives for donors to help increase the number of male donors (Andreoni and Petrie, 2004; Karlan and McConnell, 2014; IUPUI Women's Philanthropy Institute, 2015). However, this must be done in a manner that does not decrease intrinsic motivations to donate with a crowding out of pro-social behavior by external motivations (Shreedhar & Mourato, 2019). This area may also provide opportunity for future work to explore the effect of non-pecuniary incentivizes for climate activism.

*Figure 1: Donation Amount*

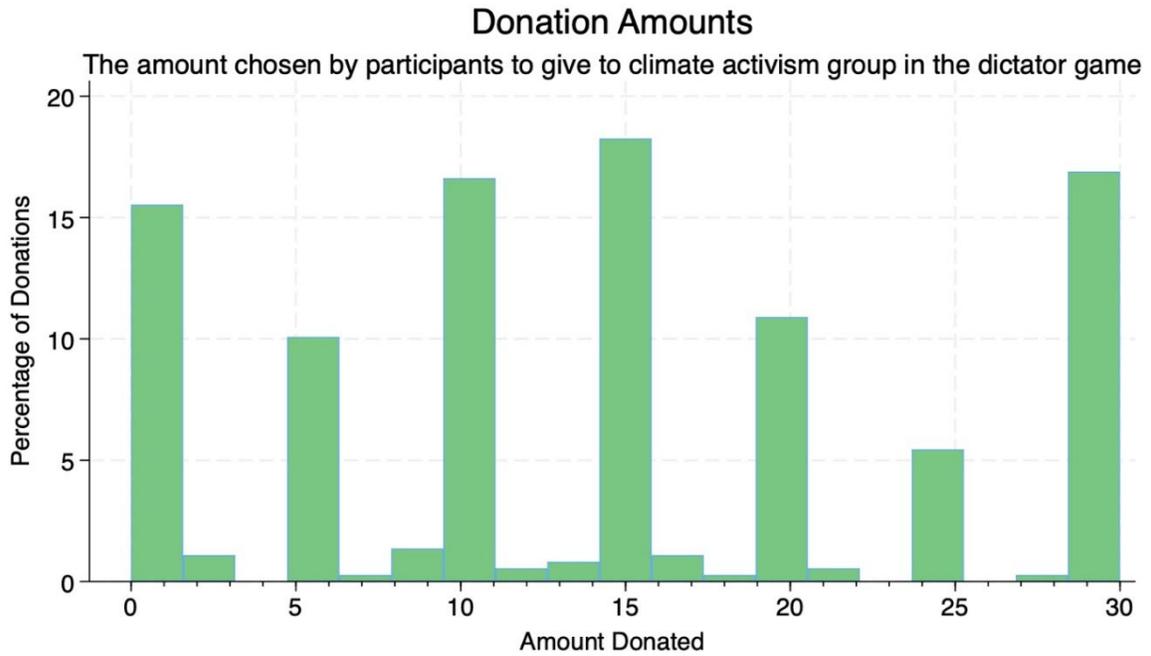

*Note*: The figure presents donation amounts, in dollars, given by participants to climate activism.

*Figure 2: Average Donations*

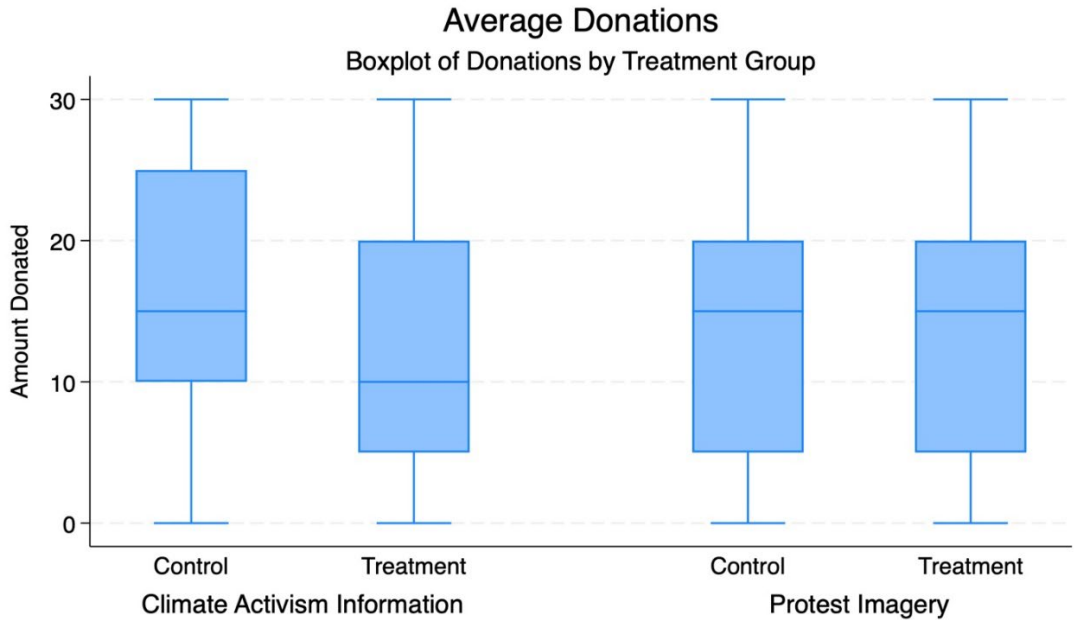

*Note*: The figure presents average donation amounts, in dollars, given by participants to climate activism, based on their random assignment to treatment or control for climate activism and for protest imagery.

*Figure 3: Average Donations for High Climate Concern*

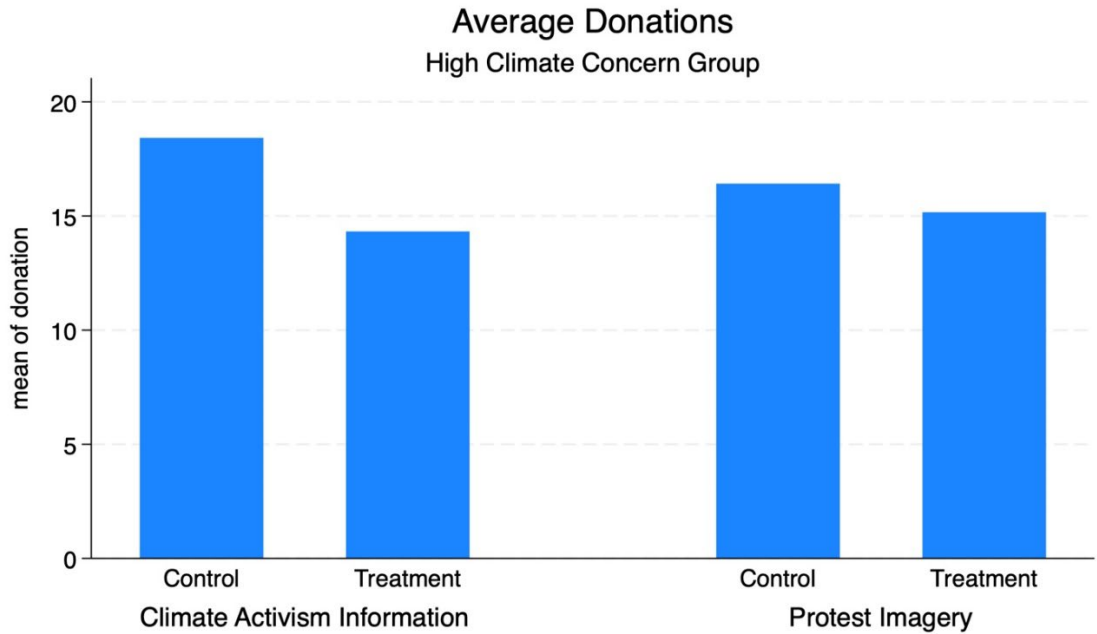

*Note*: The figure presents average donation amounts, in dollars, for the group identified as having high climate concern, based on their random assignment to treatment or control for climate activism and for protest imagery.

*Figure 4: Average Donations for Low Climate Concern*

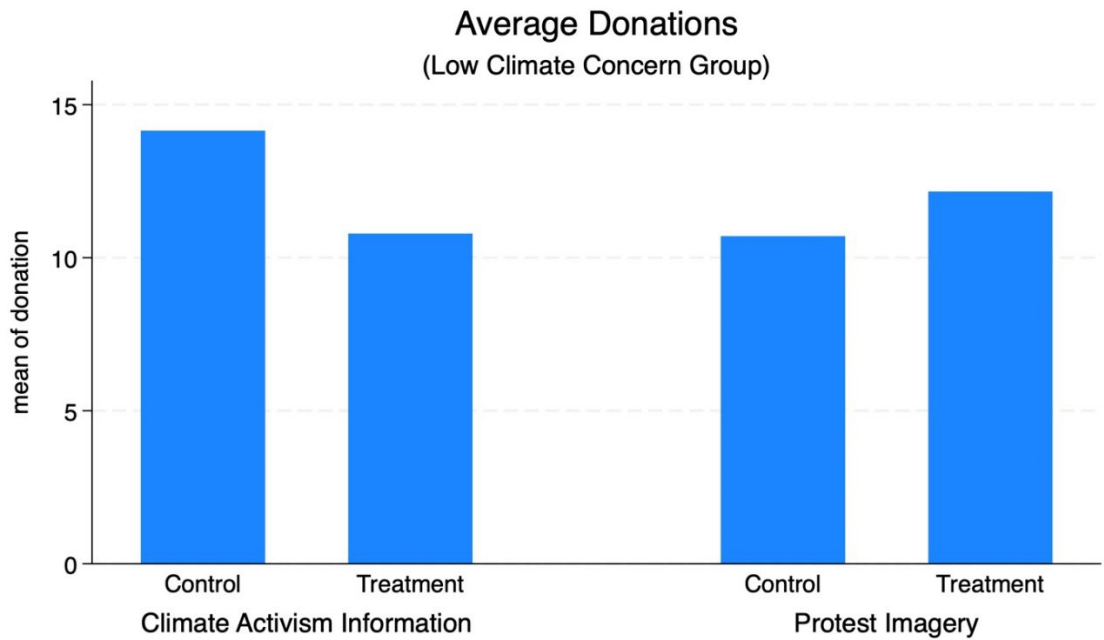

*Note*: The figure presents average donation amounts, in dollars, for the group identified as having low climate concern, based on their random assignment to treatment or control for climate activism and for protest imagery.

*Table 1: Descriptive Statistics of Variables of Interest*

|  | **Mean** | **St. Dev.** | **Min** | **Max** |
|---|---|---|---|---|
| *donation amount* | 14.29 | 9.87 | 0 | 30 |
| *post-treatment climate concern* | 16.58 | 3.36 | 5 | 21 |
| *pre-treatment climate concern* | 15.92 | 3.44 | 5 | 21 |
| *post-treatment climate change assuredness* | 10.67 | 0.74 | 5 | 11 |
| *pre-treatment climate change assuredness* | 10.24 | 1.01 | 6 | 11 |
| *climate literacy* | 16.75 | 2.31 | 10 | 22 |
| *participation in climate solutions* | 2.49 | 0.99 | 1 | 5 |
| *previous pro-social behavior* | 11.25 | 3.17 | 4 | 19 |
| *political leanings* | 38.99 | 23.85 | 0 | 100 |
| *gender* | 5.24 | 2.37 | 1 | 7 |
| *information treatment* | 0.49 | 0.50 | 0 | 1 |
| *protest treatment* | 0.50 | 0.50 | 0 | 1 |

*Note*: N = 367.

*Table 2: Tobit*

|  | Model 1 | Model 2 |
|---|---|---|
| *percent of climate concern* | 13.453*** | 11.674*** |
|  | (2.925) | (3.616) |
| *nonmale* | 5.331*** | 5.182*** |
|  | (1.502) | (1.512) |
| *protest treatment* | -3.051 | -3.066 |
|  | (1.915) | (1.916) |
| *information treatment* | -4.965** | -4.935** |
|  | (1.927) | (1.926) |
| *information protest treatment* | -3.628* | -3.526* |
|  | (1.910) | (1.907) |
| *climate change solution making* |  | 0.598 |
|  |  | (0.784) |
| *political leanings* |  | -0.027 |
|  |  | (0.031) |
| *climate literacy* |  | -0.223 |
|  |  | (0.303) |

*Note*: N = 367. *** designates statistical significance at 0.01, ** designates statistical significance at 0.05, and * designates statistical significance at 0.10.

*Table 3. Ordered Logit Regression*

|  | Donation Bin |
|---|---|
| *percent of climate concern* | 1.917*** |
|  | (0.396) |
| *nonmale* | 0.755*** |
|  | (0.213) |
| *protest treatment* | -0.375 |
|  | (0.261) |
| *information treatment* | -0.678** |
|  | (0.268) |
| *information protest treatment* | -0.491* |
|  | (0.255) |

*Note*: N = 367. *** designates statistical significance at 0.01, ** designates statistical significance at 0.05, and * designates statistical significance at 0.10.

*Table 4. Marginal Donation Effects of Each Variable on per Bin*

| Rate of Climate Concern | | |
|---|---|---|
| | 1 | -0.204*** |
| | | (0.045) |
| | 2 | -0.134*** |
| | | (0.030) |
| | 3 | -0.093*** |
| | | (0.025) |
| | 4 | 0.026* |
| | | (0.015) |
| | 5 | 0.084*** |
| | | (0.021) |
| | 6 | 0.062*** |
| | | (0.018) |
| | 7 | 0.259*** |
| | | (0.057) |
| *Non-male* | | |
| | 1 | -0.088*** |
| | | (0.029) |
| | 2 | -0.057*** |
| | | (0.017) |
| | 3 | -0.033*** |
| | | (0.009) |
| | 4 | 0.021** |
| | | (0.011) |
| | 5 | 0.038*** |
| | | (0.013) |
| | 6 | 0.025*** |
| | | (0.009) |
| | 7 | 0.092*** |
| | | (0.024) |
| *Protest Treatment* | | |
| | 1 | 0.043 |
| | | (0.032) |
| | 2 | 0.026 |
| | | (0.018) |
| | 3 | 0.016* |
| | | (0.009) |
| | 4 | -0.007 |
| | | (0.007) |
| | 5 | -0.017 |
| | | (0.012) |
| | 6 | -0.012 |

|   |   |
|---|---:|
|   | (0.008) |
| 7 | -0.048 |
|   | (0.032) |
| *Information Treatment* |   |
| *1* | 0.080** |
|   | (0.036) |
| *2* | 0.047** |
|   | (0.019) |
| *3* | 0.026*** |
|   | (0.009) |
| *4* | -0.017 |
|   | (0.011) |
| *5* | -0.032** |
|   | (0.014) |
| *6* | -0.022** |
|   | (0.009) |
| *7* | -0.083*** |
|   | (0.030) |
| *Information Protest Treatment* |   |
| *1* | 0.057* |
|   | (0.032) |
| *2* | 0.034* |
|   | (0.018) |
| *3* | 0.020** |
|   | (0.009) |
| *4* | -0.011 |
|   | (0.008) |
| *5* | -0.023* |
|   | (0.012) |
| *6* | -0.016* |
|   | (0.008) |
| *7* | -0.062** |
|   | (0.030) |

*Note*: N = 367. *** designates statistical significance at 0.01, ** designates statistical significance at 0.05, and * designates statistical significance at 0.10. Numbers 1 – 7 describe the bin into which the donation amount fell, in five dollar increments from $0 to $30.

*Table 5. Tobit Model for Climate Attitudes*

| | |
|---|---:|
| *pre-treatment climate concern* | 0.944*** |
| | (0.022) |
| *climate change solution making* | 0.138** |
| | (0.068) |
| *climate literacy* | -0.043* |
| | (0.026) |
| *nonmale* | 0.133 |
| | (0.126) |
| *political leanings* | -0.001 |
| | (0.003) |
| *protest treatment* | 0.322** |
| | (0.162) |
| *information treatment* | 0.243 |
| | (0.161) |
| *information protest treatment* | 0.221 |
| | (0.161) |

*Note*: N = 367. *** designates statistical significance at 0.01, ** designates statistical significance at 0.05, and * designates statistical significance at 0.10.

**Conflicts of Interest**

Dr. Josephson has no conflicts of interest to declare.

Ms. Gonsalves Wetherell has a friendly association with CCL and has previously attended the meetings. They had no influence over the research, research design, or this manuscript, though ultimately received $948 from this experiment.

**APPENDIX A**

*Figure A1. Visualization of Experimental Method*

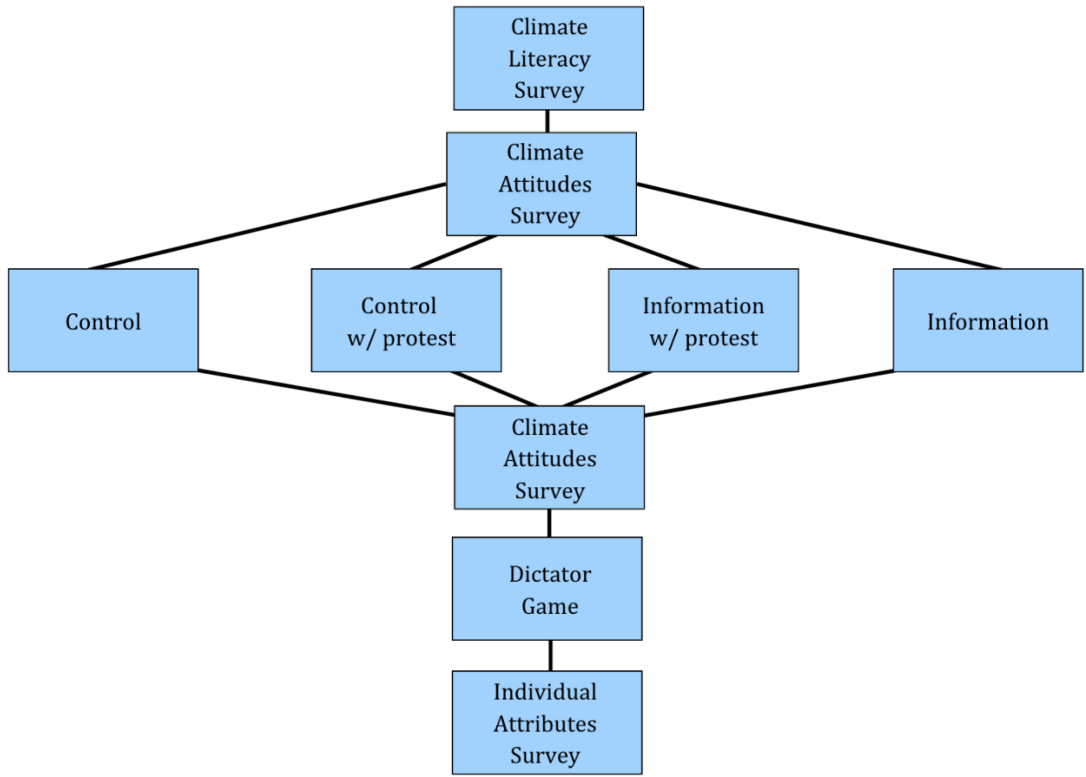

*Note*: The figure represents the design of our experimental method, the order of our experiment, and the variations in treatment that participants encountered.

*Figure A2.1: Treatment Seen for All Participants*

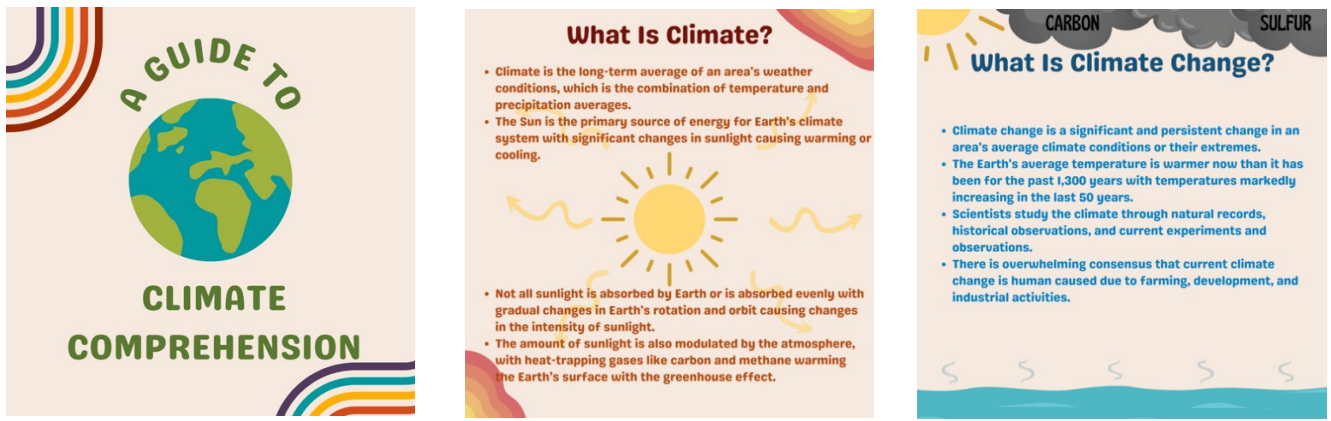

*Note*: The figures above demonstrate the three infographics seen by all participants regardless of treatment type.

*Figure A2.2. Additional Treatment Seen for No Climate Activism Information and No Protest*

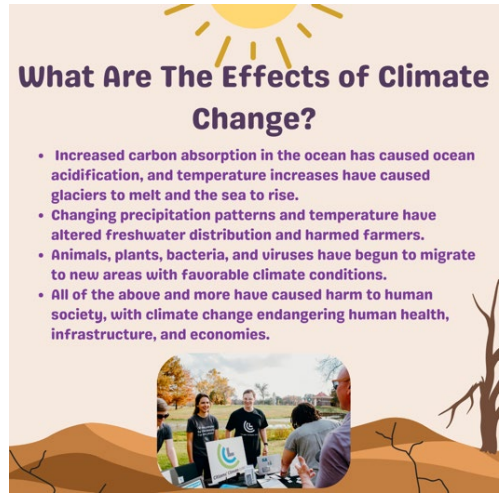

*Note*: This additional figure was seen only by those who received the complete neutral treatment of no climate activism information and no protest images.

*Figure A2.3 Additional Treatment Seen for No Climate Activism Information and Protest*

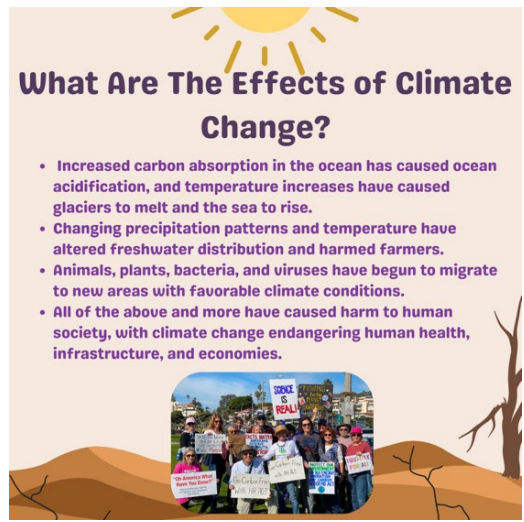

*Note*: This additional figure was seen only by those who received the neutral or no additional climate activism information treatment with the protest imagery modifier.

*Figure A2.4 Additional Treatment Seen for Climate Activism Information and No Protest*

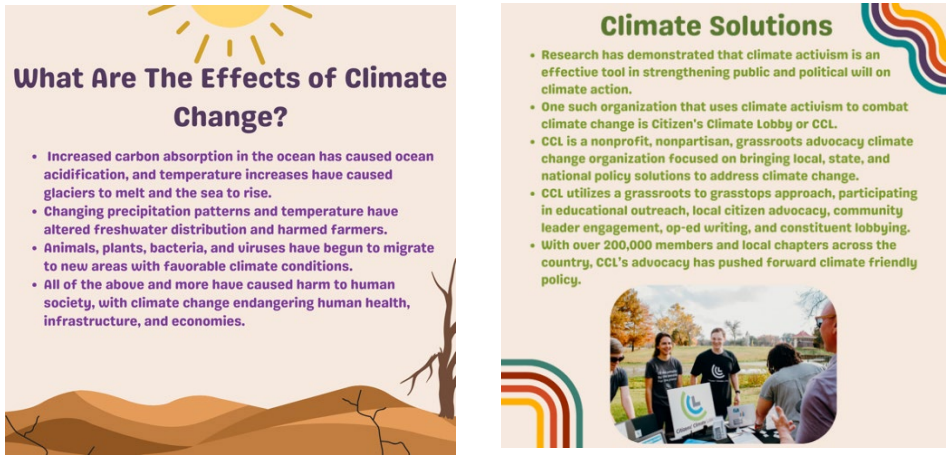

*Note*: These additional figures were seen only by those who received the additional climate activism information treatment without the protest imagery modifier.

*Figure A2.5: Additional Treatment Seen for Climate Activism Information and Protest*

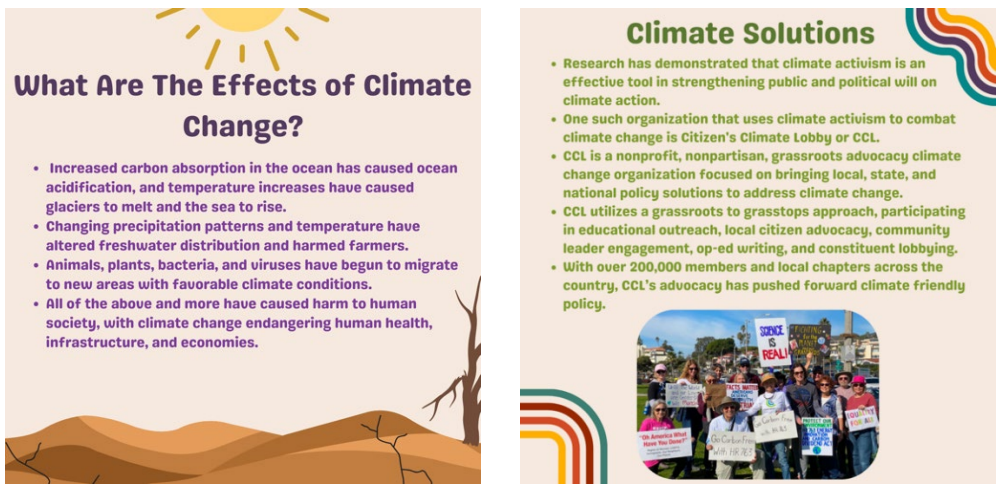

*Note*: These additional figures were seen only by those who received the additional climate activism information treatment with the protest imagery modifier.